# Polarimetry in Mössbauer spectroscopy with Synchrotron Mössbauer Source

Marina Andreeva[1,a)], Roman Baulin[1], Aleksandr Chumakov[2,3],
Tatiyana Kiseleva[1], and Rudolf Rüffer[2]

1 -   M.V. Lomonosov Moscow State University, Faculty of Physics, Moscow, Russia
2 -   ESRF-The European Synchrotron, Grenoble, France
3 -   National Research Centre "Kurchatov Institute", Moscow, Russia

a) Electronic mail: Mandreeva1@yandex.ru

Abstract

We have tested the new experimental techniques in the Mössbauer spectroscopy with Synchrotron Mössbauer Source by using the polarization analysis of the reflected radiation. In particular we have shown that the dichroic components in Mössbauer spectra create the scattering intensity with "rotated" polarization. The angular dependence for such component in the reflected signal is characterized by the peak near the critical angle of the total external reflection, and in the case of the collinear antiferromagnetic interlayer ordering the "magnetic" maxima on the reflectivity curve are formed mainly by this "rotated" polarization. The first experiment on Mössbauer reflectivity with selection of the "rotated" polarization shows the expected peak near the critical angle. The measured Mössbauer reflectivity spectra of the "rotated" polarization near the critical angle are rather different from the Mössbauer reflectivity spectra, measured without polarization analysis. They contain a smaller number of lines and can simplify the interpretation of the poorly resolved spectra. The enhanced surface sensitivity of the new technique is discussed.

I. INTRODUCTION

Interaction of light with magnetized media is characterized by specific polarization dependences. That is as well true for x-ray region of radiation. The modern synchrotrons produce or create the x-rays of any desired polarization and polarization dependent absorption or scattering near the x-ray absorption edges (XMCD, XMLD, XMND, XRMR)[1-7] has become an extremely effective methods of magnetic investigations. In the non resonant x-ray scattering the polarization analysis is used for separation of the spin and orbital magnetic moments and magnetic structure investigations [8-13].

For Mössbauer radiation the hyperfine splitting of the nuclear levels by hyperfine magnetic fields means simultaneously the energy separation by the polarization characteristics of the absorbed or reemitted quanta. The analysis of the elliptical polarization of different hyperfine transitions was done long ago[14]. The polarization dependences of Mössbauer absorption and Faraday rotation in thick samples were theoretically developed and experimentally proved in the excellent paper of Blume and Kistner[15]. In common Mössbauer spectroscopy the radioactive sources give the unpolarized single line spectrum and the polarization state of the different absorption lines has not been of special interest, but the polarization of spectral lines reveals itself in the ratio of their intensity, so the spectrum shape depends on the direction of the sample magnetization relative to the radiation propagation way[16].

The nuclear resonance experiments with synchrotron radiation have been started by using the specific way of the nuclear response registration: by detection of the time evolution of the delayed nuclear decay after prompt SR pulse[17, 18]. The hyperfine splitting of the nuclear levels leads to the quantum beats on the decay curves and the polarization of different hyperfine transitions becomes very essential: it determines the result of their interference. The waves with orthogonal polarizations does not interfere, that's why the four lines excited in magnetically split nuclear levels, when the magnetization of the sample is parallel to the SR beam of $\sigma$-polarization, give just one frequency of the quantum beats.[19] To be more precise, the coherent summation of the waves with orthogonal polarizations has no interference term in the resulting intensity[20]. However, the measurements of the scattered radiation with polarization selection immediately show the quantum beats, because the resulting linear polarization rotates with the time delay. That was splendidly demonstrated in the papers of Siddons et al.[21-22].

In the time-domain nuclear resonance spectroscopy the polarization analysis of the forward scattered radiation have been found to be very helpful for the selection of the delayed nuclear from huge prompt response at the initial delay times. The last one does not change the polarization state of the incident radiation whereas the scattering by hyperfine split nuclear levels gives the "rotated" polarization. So the polarization selection supplied the suppression of the "tale" of the prompt nonresonant radiation and allowed the monitoring of the nuclear decay from the very short delay times (after ~1 ns of their excitation)[21,23,24].

The recent development of the nuclear resonance synchrotron stations makes it possible the energy-domain Mössbauer spectroscopy with SR. In particular, the nuclear $^{57}$FeBO$_3$ monochromator (Synchrotron Mössbauer Source - SMS[25]) has been installed at the ID18 beamline[26] of the European synchrotron (ESRF). The pure nuclear reflection (333) of the iron borate crystal provides pure $\pi$-polarized radiation at 14.4 keV within a bandwidth of 8 neV. The usage of this $\pi$-polarized beam results in the new features of the Mössbauer spectra measured in the absorption or reflection geometry with SMS[27] compared with laboratory experiments. It can be expected that the polarization analysis of the scattered or reflected intensity will be even more informative for the investigation of the magnetic properties of samples. We can refer to the analogy with the polarized neutron reflectivity, in which the curves with spin-flip neutrons are very informative. In the nonresonant magnetic x-ray scattering the polarization analysis is also effectively used.

In this work, the first results demonstrating the peculiarities of the nuclear resonant reflectivity (NRR) with SMS supplemented by the polarization analysis of the reflected intensity are presented. We show the differences of the nuclear resonant reflectivity curves with unchanged and with "rotated" polarization and explain the results using the x-ray standing wave approach. The practical significance of this new development for the complicated spectra treatment is also discussed.

II. THEORY

We start from the model calculations of the NRR angular curves and Mössbauer reflectivity spectra at several grazing angles with selection of the reflected radiation by the polarization state – Fig. 1. The general theory of NRR has been thoroughly developed in several papers[28-30], it gives the matrix of the reflectivity amplitudes in $\sigma$- and $\pi$- polarization orts, so the reflected intensity for any polarization of the incident and reflected waves can be easily calculated for arbitrary multilayer structure (the program pack REFTIM and RESPC is placed on the ESRF webside[31]). In the energy domain case the angular dependence of the NRR can be calculated for selected energy in the resonant spectrum range or as the integral over Mössbauer reflectivity spectra at each grazing angle $\theta$ (the results in Fig. 1a-d are obtained by this integrated mode, and it corresponds to the experimental procedure with SMS). Calculations for Fig. 1 have been done for the [$^{57}$Fe(0.8nm)/Cr(2nm)]$_{30}$ multilayer, the hyperfine magnetic field $B_{hf}$=33 T with distribution $\Delta B_{hf}$=1 T is supposed for $^{57}$Fe nuclei. Note that 14.4 keV Mössbauer transition is of magnetic dipole M1 type, so the magnetic field of radiation H$^{rad}$ interacts with $^{57}$Fe nuclei. In the case of the $\pi$-polarized radiation from SMS we consider the radiation field vector H$^{rad}$, which is lying in the sample surface.

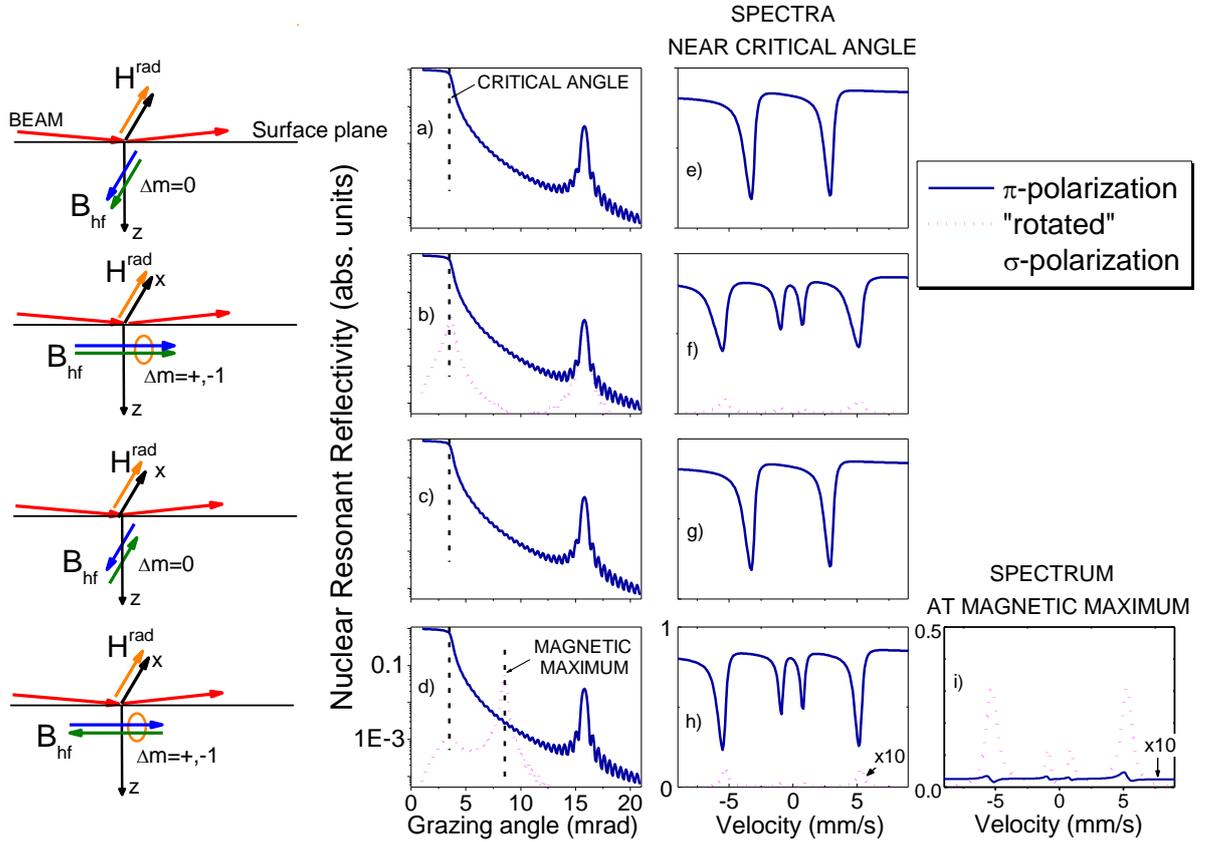

FIG. 1. Model calculation of the NRR curves (left graph column) and Mössbauer spectra of reflectivity at the critical angle and in the "magnetic" maximum (middle and right graph columns correspondingly) for $\pi$- polarized SMS radiation and for different cases of the ferromagnetic and antiferromagnetic coupling between adjacent $^{57}$Fe layers, schematically drawn on the left. The hyperfine transitions $\Delta m=0,\pm 1$ excited in parallel and perpendicular magnetization orientations relative SMS beam are also marked. Solid (blue on-line) lines represent the $\pi\rightarrow\pi$ reflectivity and dash (purple on-line) lines represent $\pi\rightarrow\sigma$ reflectivity.

The results of these simple calculations show some unexpected features. The shapes of the NRR angular reflectivity curves for the $\pi\rightarrow\pi$ and $\pi\rightarrow\sigma$ scattering are found very different. The NRR $\pi\rightarrow\pi$ angular curve approaches 1 when $\theta\rightarrow 0$. This behavior directly follow from usual Fesnal law, and had been observed in the first paper devoted to the Mössbauer reflectivity[32]. Notice that in the time-domain experiments the shape of the NRR curve is different[33], because the energy modulation of NRR amplitude in resonant region diminishing up to zero at $\theta \to 0$ and their Fourier transform (which is necessary in transition to the time-domain) leads to 0 at $\theta \to 0$. The NRR angular curves with "rotated" σ-polarization $I_{\pi \to \sigma}(\theta)$ as well lead to 0 if $\theta \to 0$ (Fig. 1b,d). Besides they are characterized by the sharp peak at the critical angle of the total external reflection. For the case, when the antiferromagnetic interlayer coupling makes the magnetic period of the structure twice larger the chemical period, the additional Bragg peak ("magnetic" maximum) appears, but only on the reflectivity curve with the "rotated" polarization (Fig. 1d). The appearance of the "rotated" polarization in the reflected signal takes place only when $B_{hf}$ has projections on the beam direction and the resonant transitions with $\Delta m = \pm 1$ are characterized by the circular polarization of the reemitted radiation. That's why not six but only four lines with $\Delta m = \pm 1$ are presented in the Mössbauer reflectivity spectrum with "rotated" polarization. Practically only the "rotated" σ-polarization contributes to the "magnetic" maximum (provided that there are no canted $B_{hf}$ in the adjacent $^{57}$Fe layers[27]).

The peak at the critical angle has the similar origin as in the case of the time-domain NRR angular curve. There are several explanations of this feature for the time-domain measurements[33-35], we prefer to follow the approach in which the x-ray standing waves, created by the prompt pulse, are responsible for the resonant nuclei excitation[34-35]. For NRR angular curve in the energy-domain the same standing wave mechanism can be enabled. If the dichroic contribution to the standing wave formation inside the sample is small, the NRR angular curve for the "rotated" polarization $I_{\pi \to \sigma}(\theta)$ can be calculated by the expression:

$$I_{\pi \to \sigma}(\theta) = \left| \frac{i\pi}{\lambda \sin\theta} \int \chi^{\pi \to \sigma}(z) E_\pi^{\ 2}(\theta, z)\, dz \right|^2, \qquad (1)$$

where $\chi^{\pi \to \sigma}$ is the off-diagonal component of the susceptibility matrix in the σ,π-polarization orts (dichroic component). The expression (1) contains the full radiation field of the dominant π-polarization $E_\pi$ in the 4$^{th}$ power ("squared standing wave"), and it is well known in the reflectivity theory that the x-ray standing wave near the sample surface has a maximum at the critical angle[36,37]. It explains the peak near the critical angle on the NRR curve for the "rotated" σ-polarization. The direct comparison between the results of calculations by the formula (1) with the exact calculations[31] shows the excellent agreement provided that the resonant contribution to the susceptibility $\chi^{\pi\to\sigma}$ can be considered as a small correction.

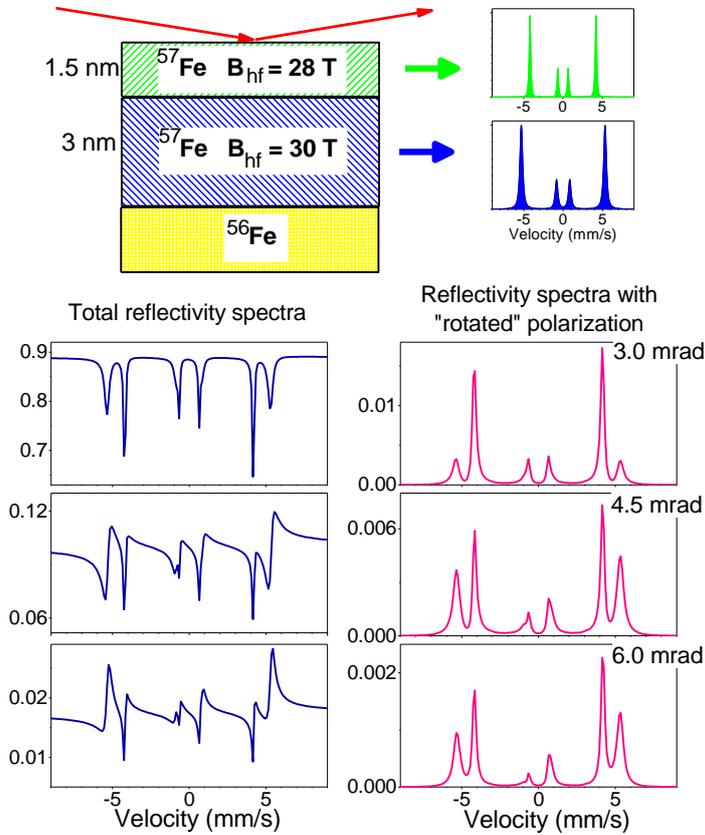

FIG. 2. *Model calculation of the Mössbauer reflectivity spectra, normalized to the incident intensity of π-polarization, without the polarization selection (π→π+σ) and for selected (π→σ) "rotated" polarization (right graph column), calculated for three grazing angles in vicinity of the the critical one. $B_{hf}$ along the beam direction. The used model is schematically drawn on the top.*

The ideology, appealing to the standing wave influence on the reflectivity with "rotated" polarization, have the important consequence on the practical importance of such measurements with polarization selection. Notably, the analogy with the secondary radiation measurements (even enhanced by the "squared standing wave") points to the much more effective depth selectivity of this dichroic signal comparing with the common measurements of the reflectivity. The model calculations illustrate the enhanced surface sensitivity of the signal with the "rotated" σ-polarization (Fig. 2). In order to distinguish the thin top layer (1.5 nm $^{57}$Fe) from the bottom one we supposed that in the top layer (1.5 nm thickness) the resonant spectrum corresponds to the $B_{hf}$ = 28 T, and for the deeper $^{57}$Fe layer (3 nm thickness) $B_{hf}$ = 30 T. The calculated Mössbauer reflectivity spectra at several grazing angles near the critical one without and with the "rotated" σ-polarization selection clearly show that the spectra with "rotated" σ-polarization contain more intense contribution to the spectrum from the 1.5 nm top layer than ordinary Mössbauer reflectivity spectra.

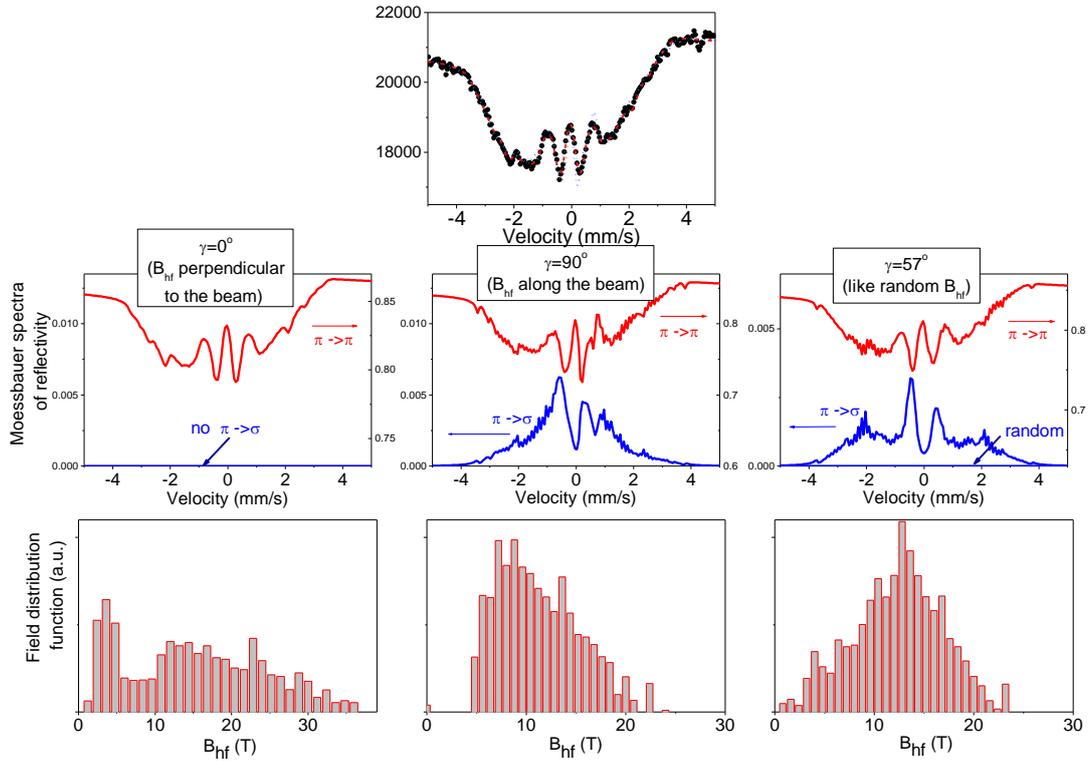

FIG. 3. *Mössbauer reflectivity spectrum from the Ref.[38], measured in remanence from [Fe(0.12 nm)/Cr(1.05 nm)]\*30 sample, (top); symbols are the experimental data, three solid (practically undesignable) lines are the fit results for three different field orientations: for $B_{hf}$ in the surface plane and perpendicular to the beam (azimuth angle γ=0°), $B_{hf}$ parallel to the beam (γ=90°)) and $B_{hf}$ in the surface plane with γ=57° (at this angle the Mössbauer reflectivity spectrum without polarization analysis has the same shape as in the case of the random in space $B_{hf}$ orientations) and corresponding different field distributions $P(B_{hf})$ for these three cases (bottom), giving the same fit result to the experimental spectrum on the top. Mössbauer reflectivity spectra with polarization selection, for these models are shown in the middle: dash lines (red on-line) for the "nonrotated" (π→π) polarization, dot lines (blue on-line) for the "rotated" (π→σ) polarization of the reflected radiation. In the cases of the $B_{hf}$, distributed randomly in space, and if $B_{hf}$ is perpendicular to the scattering plane, the "rotated" polarization is absent.*

The next advantage of the measurements with polarization selection is illustrated by Fig. 3. This selection simultaneously means the decreasing of the number of lines in Mossbaur reflectivity spectra. It can be very useful for the interpretation of the poorly resolved complicated spectra. The experimental Mossbaur reflectivity spectrum[38], presented on the top of Fig. 3, can be fitted by at least three different models with completely different field orientation and distribution. The spectra with "rotated" polarization are quite different for these three cases (Fig. 3). So, the true picture of the hyperfine field distribution and orientation can be recovered if we could perform the polarization analysis of the reflectivity spectra.

## III. EXPERIMENTAL RESULTS AND DISCUSSION

The experiment was performed at the ID18 beamline[26] of the ESRF. The storage ring was operated in multi-bunch mode with a nominal storage ring current of 200 mA. The energy bandwidth of radiation was first reduced down to 2.1 eV by the high-heat-load monochromator[39] adjusted to the 14.4125 keV energy of the nuclear resonant transition of the $^{57}$Fe isotope. Then X-rays were collimated by the compound refractive lenses down to the angular divergence of a few $\mu$rad. The high-resolution monochromator (HRM) decreases the energy bandwidth of the beam further to ~15 meV. Final monochromatization down to the energy bandwidth of ~8 neV was achieved with SMS[25] and the sweep through the energy range of a Mössbauer resonant spectrum (about ±0.5 $\mu$eV) was achieved using the Doppler velocity scan. Radiation from the SMS was focused down to the beam size of 8×10 $\mu m^2$ using the Kirkpatrick-Baez multilayer mirror system. The intensity of the X-ray beam incident on the sample was ~$10^4$ photons/s. The selection of the "rotated" σ-polarization was performed by Si channel-cut crystal (two (840) reflection with $\theta_B$ =45.1$^o$ for 14.4 keV radiation) – Fig. 4.

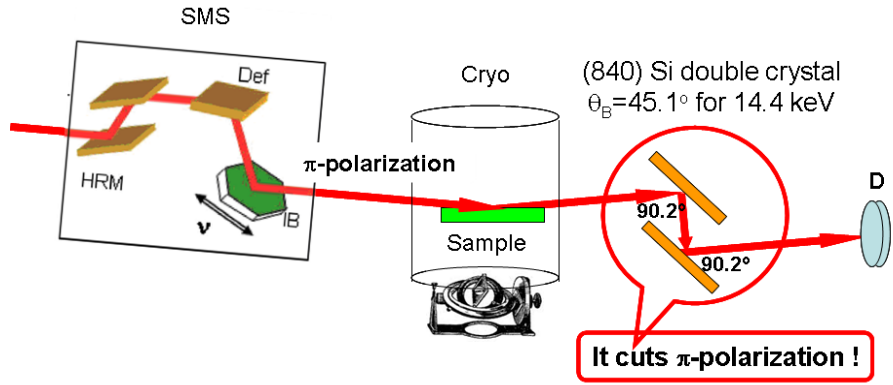

FIG. 4. *Experimental set-up for the Mössbauer relectivity measurements with selection of the "rotated" σ-polarization.*

Several [$^{57}$Fe/Cr]$_{30}$ samples were mounted in the cassette holder of the He-exchange gas superconducting cryo-magnetic system and the measurements were performed at helium temperature (4 K). To ensure the ferromagnetic alignment of the magnetization in the $^{57}$Fe layers and to increase the NRR reflectivity with the "rotated" polarization near the critical angle the external magnetic field (5 T) was applied along the beam direction.

The measurements of the reflectivity with "rotated" polarization were exaggerated by two circumstances. The double-reflection from the channel-cut Si polarization-analyzer cuts ~100% of the π-polarized radiation but its efficiency for the σ-polarized radiation is only 4 % (96 % we lose), besides the analyzer has the very small angular acceptance (~10"). This had a crashing effect due to the unexpected bad quality of the sample surfaces. Some of them were curved and others had mosaic-like surface. The angular divergence of the reflected radiation (as it was revealed by the angular scan with the analyzer) was ~200". So the total efficiency of the detection of the "rotated" polarization intensity was very low. The measurements at the weak "magnetic" maxima from the samples with antiferromagnetic interlayer coupling were prohibitive, so it was decided to measure near the critical angle applying the strong magnetic field (5 T) along the beam direction to ensure the ferromagnetic alignment in our samples.

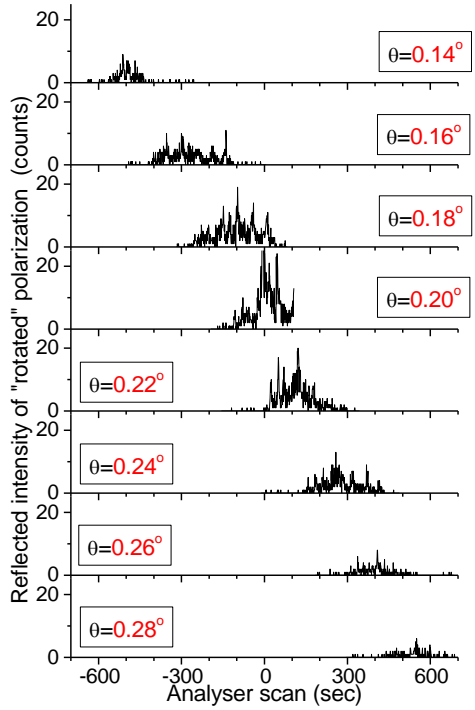

FIG. 5. *Angular scans of the reflectivity divergence from [$^{57}$Fe(0.8 nm)/Cr(1.05 nm)]$_{30}$ sample measured at different grazing angles with polarization analyzer tuned for the "rotated" polarization ($\pi \to \sigma$).*

The scans, presented in Fig. 5, were obtained with the line width of the SMS deliberately enlarged up to 5 mm/s (it is possible by small deviation of the angle in the region of the diffraction maximum from $^{57}$FeO$_3$ – for details see Ref.[25]) in order to overlap as much nuclear resonant lines as possible and to increase the NRR intensity. The NRR curve for the "rotated" $\sigma$-polarization (Fig. 6) was obtained as an integral over these angular scans at each grazing angle.

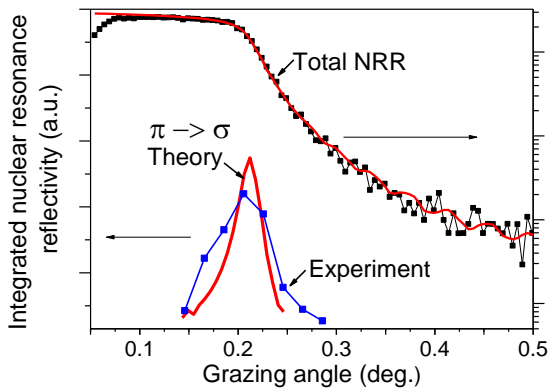

FIG. 6. *NRR angular curves measured without polarization selection (logarithmic right scale) and with the selection of the $\sigma$-polarized reflectivity (left scale), obtained as the integral over scans, presented in FIG. 5.*

The obtained NRR curve with "rotated" σ-polarization in Fig. 6 shows the expected peak near the critical angle, the reason of its appearance has been explained in the previous section. On the NRR curve measured without polarization selection the plateau below the critical angle is well seen.

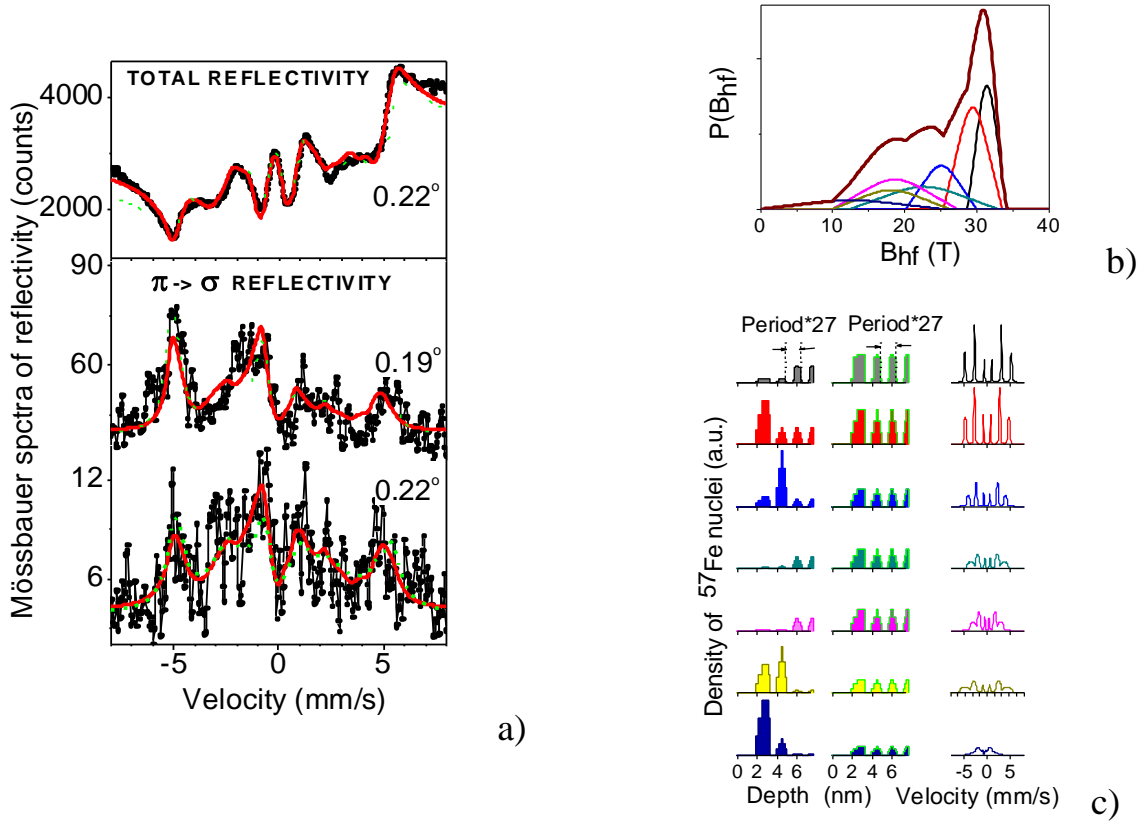

Fig. 7. (a) *Mössbauer reflectivity spectra measured from [$^{57}$Fe(0.8 nm)/Cr(1.05 nm)]$_{30}$ multilayer without polarization analysis at the critical angle (θ=0.22°) and with selection of the "rotated" polarization at the two angles in vicinity of the critical angle; (b) Hyperfine field distribution, obtained by the fit (solid, red on-line lines in (a)); (c) Depth distribution of the nuclei for each of $B_{hf}^{(i)}$(i=1,7), obtained by the fit (left column), the suggested initially homogeneous depth distribution of $B_{hf}^{(i)}$ in all $^{57}$Fe layers, giving the calculated reflectivity spectra in (a), drawn by the dashed (green on-line) lines (middle column); resonant spectra, corresponding to every $B_{hf}^{(i)}$(right column).*

Mössbauer reflectivity spectra with the "rotated" σ-polarization were measured at the two grazing angels in vicinity of the critical angle (θ=0.19° and

θ=0.22º) – Fig. 7a. In spite of the bad statistics, stipulated by the huge loss of the intensity during polarization selection with 840-Si double reflections, we can recognize some essential features. The resonant lines are presented in these spectra as peaks, as it was predicted by the theory (see Figs.1b and 1f). The spectra with the "rotated" σ-polarization essentially differ from the spectrum measured without polarization analysis, the lines in which have dispersive-like shape and which is also presented in Fig. 7.

We were surprised by the asymmetry and shape difference between the spectra at θ=0.19º and θ=0.22º. At the beginning we supposed that these effects are due to a small contribution of the "non-rotated" reflectivity. But the model calculations did not confirm such assumption. Finally the explanation was found.

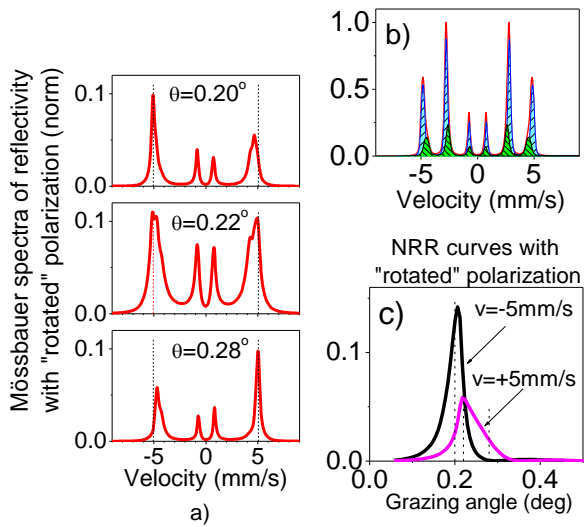

FIG. 8. *(a) Model calculations of the shape variation of the Mössbauer reflectivity spectrum with "rotated" polarization caused by the small change of the grazing angle in vicinity of the critical angle. Dashed vertical lines mark the energies for which the NRR curves in (c) are calculated. Calculations have been performed for the model of 10nm $^{57}$Fe layer on $^{56}$Fe substrate in which two hyperfine fields $B_{hf}$ = 28 T and 30 T are presented, resonant spectra for them are drown in b). (c) NNR angular curves with "rotated" polarization, calculated for the 1$^{st}$ (v=-5mm/s) and 6$^{th}$ lines (v=+5mm/s) of the $B_{hf}$ =30 T sextet. Dashed vertical lines mark the angles, for which the spectra in (a) are calculated.*

The effects are explained by the refraction effect. The NNR curves with "rotated" σ-polarization calculated for the photon energies corresponding to the two outer lines in Mössbauer sextet (actually quartet) show some shift of the critical angle peak (Fig. 8c) which is caused by the refraction initiated by the contribution to the susceptibility from the lines of the additional multiplet of a smaller splitting (in the experiment we have even more than two smaller $B_{hf}^{(i)}$). So a small angular shift in vicinity of the critical angle brings sequentially to the most intense reflectivity one line after another one in Mössbauer reflectivity spectrum with "rotated" polarization. The same effect was observed at the spectra measured in the "magnetic" maximum for the same sample[40] (which is basically $\pi \rightarrow \sigma$, see the model NRR and spectrum in Fig. 1d and 1i).

The shapes of the spectra in Fig. 7 can be more or less reproduced by the fit performed with REFSPC program package[31]. All three spectra were treated within the same model. We have used seven hyperfine fields $B_{hf}^{(i)}$, i=1,7, their fitted parameters and distribution probabilities are shown in Fig. 7b. The obtained depth distributions for each $B_{hf}^{(i)}$ are presented in the first column of Fig. 7c. We can see that the hyperfine fields in the top two $^{57}$Fe layers are essentially different from the following periodic layers: in average the hyperfine splitting is smaller in the top layers and the portion of the low field contributions is enlarged. Assuming the same depth distribution in the top layers as in the whole periodic part we get the theoretical spectra, presented by dashed lines in Fig. 7a, which show worse correspondence with the experimental spectra. This result confirms the enhanced depth selectivity of the spectra with "rotated" polarization.

IV. CONCLUSIONS

We have performed the first experiment in which the nuclear resonant reflectivity angular curve, measured with the "rotated" polarization in the energy-domain mode, shows the peak near the critical angle of the total reflection. The enhanced surface sensitivity of the Mössbauer spectra of reflectivity with "rotated" polarization has been supposed and confirmed.

We hope that in future the technique for the polarization analysis will be improved and it becomes the effective tool for the solving the uncertainties in the interpretation of the complex Mössbauer spectra.

ACKNOWLEDGMENTS

This work was supported by the Russian Foundation of the Basic Research through the Grants No. 15-02-1502 and 16-02-00887-a. The authors are thankful to Ya.A. Babanov, D. A. Ponomarev and M. A. Milyaev for the assigned samples.